\documentclass[12pt]{article}
\usepackage{aas_macros}
\usepackage{amssymb}
\usepackage{amsmath}
\usepackage{mathtools}
\usepackage{array}
\usepackage[dvipdfmx]{graphicx}
\usepackage{color}
\usepackage{cite}
\usepackage{url}
\usepackage{bm}
\usepackage{float}
\usepackage[colorlinks,citecolor=blue]{hyperref}
\renewcommand{\thefootnote}{\#\arabic{footnote}}

\renewcommand{\thefootnote}{\fnsymbol{footnote}}
\def\thefootnote{\fnsymbol{footnote}}

\makeatletter

\@addtoreset{equation}{section}
\makeatother

\newcommand{\mrm}{\mathrm}

\newcommand{\mcal}{\mathcal}

\newcommand{\si}{\sigma}

\newcommand{\om}{\omega}

\newcommand{\De}{\Delta}
\newcommand{\ep}{\epsilon}

\begin{document}

\begin{titlepage}

\begin{center}

\vskip .75in

{\Large \bf Supermassive primordial black holes: \vspace{2mm} \\ a view from clustering of quasars at $z\sim 6$}

\vskip .75in

{\large
Takumi~Shinohara$\,^1$, Wanqiu~He$\,^2$,  Yoshiki~Matsuoka$\,^3$, \\[4pt]
Tohru~Nagao$\,^3$, Teruaki~Suyama$\,^4$,  and  Tomo~Takahashi$\,^5$ 
}

\vskip 0.25in

{\em
$^{1}$Graduate School of Science and Engineering, Saga University, Saga 840-8502, Japan
\vspace{2mm} \\
$^{2}$National Astronomical Observatory of Japan,  2-21-1 Osawa,  Mitaka, \\  Tokyo  181-8588, Japan
\vspace{2mm} \\
$^{3}$Research Center for Space and Cosmic Evolution, Ehime University, Matsuyama, Ehime 790-8577, Japan 
\vspace{2mm} \\
$^{4}$Department of Physics, Tokyo Institute of Technology, 2-12-1 Ookayama, Meguro-ku,
Tokyo 152-8551, Japan
\vspace{2mm} \\
$^{5}$Department of Physics, Saga University, Saga 840-8502, Japan
}

\end{center}
\vskip .5in

\begin{abstract}

We investigate a scenario where primordial black holes (PBHs) can be the progenitors of supermassive black holes (SMBHs) observed at $z\sim6$. To this end, we carried out clustering analysis using a sample of 81 quasars at $5.88 <z<6.49$, which is constructed in Subaru High-$z$ Exploration of Low-Luminosity Quasars (SHELLQs) project, and 11 quasars in the same redshift range selected from the literature. The resulting angular auto-correlation function (ACF) can be fitted to a power-law form of $\omega_\theta = 0.045^{+0.114}_{-0.106}~\theta^{-0.8}$ over a scale of $0.2$ -- $10$ degrees. We compare the ACF of the quasars to that predicted for the PBH model at $z\sim 6$ and found that  such a scenario is excluded for a broad range of parameter space, from which we can conclude that a scenario with PBHs as SMBHs is not viable. We also discuss a model in which SMBHs at $z \sim 6$ originate from the direct collapse of PBH clumps and argue that the observed ACF excludes such a scenario in the context of our PBH model.

\end{abstract}

\end{titlepage}

\renewcommand{\thepage}{\arabic{page}}
\setcounter{page}{1}
\renewcommand{\thefootnote}{\#\arabic{footnote}}
\setcounter{footnote}{0}

\section{Introduction \label{sec:intro}}

Observational wide-field optical and near-infrared surveys such as Sloan Digital Sky Survey (SDSS) \cite{2001AJ....122.2833F,2016ApJ...833..222J}, Subaru High-$z$ Exploration of Low-Luminosity Quasars (SHELLQs) \cite{2016ApJ...828...26M,2019ApJ...872L...2M,2022ApJS..259...18M}, Canada-France-Hawaii Telescope Legacy Survey (CFHTLS) \cite{2007AJ....134.2435W,2010AJ....139..906W}, Panoramic Survey Telescope and Rapid Response System 1 (PanSTARRS1) \cite{2016ApJS..227...11B}, and United Kingdom Infrared Telescope Infrared Deep Sky Survey (UKIDSS) \cite{2011Natur.474..616M} have discovered quasars powered by supermassive black holes (SMBHs) at redshift around and higher than 6. The virial black hole mass estimated from their single-epoch spectra suggest that the mass of SMBHs in such high-$z$ quasars had been already grown up to $\sim 10^8 - 10^{10} M_\odot$ even when the age of the Universe was less than 1 Gyr (e.g., \cite{2015Natur.518..512W,2019ApJ...880...77O,2022ApJ...941..106F}). However, the formation of SMBHs at high redshift from astrophysical processes is a challenging task and a lot of efforts have been made to understand the mechanism (for a review, see, e.g., \cite{Inayoshi:2019fun}). 

Another avenue to explain SMBHs at high-$z$ is primordial black holes (PBHs) \cite{Hawking:1971ei,Carr:1974nx,Carr:1975qj} as a progenitor, which has been attracting attention in cosmology. 
Indeed PBHs have been a target of intense study since it can explain dark matter of the Universe for some mass range \cite{Carr:2020gox} and the gravitational wave signals detected by LIGO \cite{Bird:2016dcv,Clesse:2016vqa,Sasaki:2016jop}. 
An advantage of considering PBHs as the origin of SMBHs is that the initial mass of PBHs is 
specified by the characteristic comoving wavenumber which is 
a free parameter characterizing the primordial power spectrum  
and it is possible to produce PBHs whose initial mass is in the mass range $10^8 M_\odot - 10^{10} M_\odot$ by simply adopting appropriate value
of the comoving wavenumber (see Eq.~(\ref{k-MPBH})),
which is in sharp contrast with Pop-III scenario for which additional astrophysical process to achieve efficient mass growth from initial masses $\lesssim 10^3 M_\odot$ is needed \cite{Hirano:2013lba, Johnson:2006gd, Smith_2018}.
On the other hand, a non-trivial issue is that a significant enhancement of the primordial power spectrum from the standard power-law form with the spectral index $n_s \approx 0.96$ 
must occur at small scales to realize the ${\cal O}(1)$ amplitude of the primordial curvature perturbations required for the PBH formation,
as it is argued in \cite{Duechting:2004dk}.
A possibility that PBHs play a role of SMBHs has been discussed 
in the context of some concrete models in \cite{Kawasaki:2012kn,Nakama:2016kfq,Hasegawa:2017jtk,Kawasaki:2019iis,Kitajima:2020kig}.  

Although the direct formation of the SMBHs is an appealing feature of the PBH scenario,
it is logically possible that PBHs are formed with their initial masses $\lesssim 10^3 M_\odot$ and grow to SMBHs until the redshifts of observations.
Yet, there is no strong motivation to consider this possibility since Pop-III stars can be such initial seed BHs.
Thus, except the last paragraph in Sec.~\ref{sec:constraint}, throughout this paper, 
we assume that initial PBH masses are in the mass range $ 10^4 M_\odot - 10^{10} M_\odot$\footnote{
For the PBH mass of $\lesssim 10^8 M_\odot $, some astrophysical process such as accretion is required so that  PBHs grows to obtain observed masses of SMBHs. 
}.

It is known that PBH formation in the mass range $ 10^4 M_\odot - 10^{10} M_\odot$ contradicts 
with non-detection of the spectral distortion of cosmic microwave background (CMB) 
if such PBHs are produced from Gaussian primordial fluctuations \cite{Kohri:2014lza}. 
Therefore one needs to invoke a highly non-Gaussian perturbation to successfully produce PBHs as SMBHs
without conflicting with this constraint.
One natural path to realize such a scenario is to introduce a light spectator field whose fluctuations produced
during inflation source the non-Gaussian curvature perturbation at late epoch\footnote{
Recent work \cite{Deng:2017uwc} proposed a new scenario based on multi-field inflation models 
where bubbles formed through the tunneling process during inflation can lead to PBHs, 
which does not make use of the standard quantum fluctuations of fields produced during inflation. 
In this scenario, CMB spectral distortion is not induced. 
Furthermore, if the isocurvature modes are heavy, the tunneling rate will depend only on the adiabatic direction. 
In such a case, the spatial distribution of PBHs will obey the Poisson distribution. \label{f-bub}
}.
This idea has been framed in concrete models in \cite{Nakama:2016kfq,Hasegawa:2017jtk,Kawasaki:2019iis,Kitajima:2020kig}. Recently, in \cite{Shinohara:2021psq}, three of the present authors have shown that the distribution of PBHs is highly clustered 
in such models by explicitly demonstrating that the computed angular correlation function of PBHs becomes large over
a wide range of angles (see also \cite{Kawasaki:2021zir}). 
Since the essence of the clustering comes from the modulation of the local amplitude of the perturbation
(at the scales corresponding to the PBH formation)
caused by the superposition of the long-wavelength modes, 
the strong clustering of PBHs is a generic feature in the light spectator field models\footnote{
In general, when short-wavelength perturbations are modulated by the long-wavelength perturbations
by the presence of local non-Gaussianity,
the resultant PBHs are clustered \cite{Tada:2015noa, Young:2015kda, Suyama:2019cst}.
}. 

Although it was suggested in \cite{Shinohara:2021psq} that large amplitude of the angular correlation function
is likely to be incompatible with the observed quasars,
a definite conclusion was not drawn because the angular correlation function (ACF) predicted in the model was not confronted with
observational data.
This paper takes forward by comparing the angular correlation function computed in \cite{Shinohara:2021psq} with 
actual observational data as a critical test for the PBH scenario as SMBHs.  
To this end, we derive the angular correlation function for quasars at $ 5.88 < z < 6.49 $ discovered in SHELLQs project, and additional quasars in the same redshift range detected in the literature \cite{Matsuoka:2021jlr}.
Then we compare the ACF of quasars with that of PBHs, which is predicted  \cite{Shinohara:2021psq}, and show that most parameter space are likely to be excluded by the observational result.
In particular, the case where the mass of the spectator field is much smaller than the Hubble parameter
during inflation is rejected. 
Therefore, as we argue in this paper, we can conclude that PBH models based on the light spectator field 
as a progenitor of SMBHs are not viable.

This paper is organized as follows. In Section~\ref{sec:PBH_SMBH}, we briefly summarize our model to derive the clustering of PBHs at $z \sim 6$. In Section~\ref{sec:analysis}, we introduce the quasar sample used for clustering analysis in this study, and evaluate the clustering of quasars at $z \sim 6$ via the ACF. Then the comparison between the two ACFs and the constraints on model parameter space are discussed in Section~\ref{sec:constraint}. We give conclusions in the final Section.

\section{PBHs as SMBHs \label{sec:PBH_SMBH}}

In this section, based on Ref.~\cite{Shinohara:2021psq},
we first give a brief overview of the scenario which explains the existence of
the SMBHs at redshifts around $z=6$ as a result of formation of PBHs
in the very early Universe.
After that, we briefly give a formalism to evaluate the angular correlation function of PBHs
which we compare with observational data of quasars at $z\sim6$.

A PBH is formed when the primordial density perturbation which has a density contrast of
${\cal O}(1)$ amplitude re-enters the Hubble horizon \cite{Carr:1974nx}. 
Crudely speaking, the initial PBH mass $M_{\rm PBH}$ is equal to
the horizon mass evaluated at the time of the horizon reentry,
which gives a relation between the (comoving) wavenumber $k$ 
of the perturbation and $M_{\rm PBH}$ as
\begin{equation}
\label{k-MPBH}
k \sim 100\,\mrm{Mpc}^{-1} \biggl(\frac{M_\mrm{PBH}}{10^{10} M_\odot}\biggr)^{-1/2}.
\end{equation}
Such a large-amplitude perturbation must be very rare, 
otherwise the SMBHs are over-produced and the Universe would be dominated
soon after their formation\footnote{
The formation time of PBH with its mass $M_{\rm PBH}$ is 
$t\sim 10^5~{\rm s}~\left( \frac{M_{\rm PBH}}{10^{10} M_\odot} \right)$.
}, which is not compatible with reality.
If the primordial perturbations of the mode with wave number $k$ corresponding to 
$10^4 M_\odot \lesssim M_{\rm PBH} \lesssim 10^{10} M_\odot$ are Gaussian/nearly Gaussian,
it is known that such perturbations lead to a sizable CMB spectral distortion incompatible
with the non-detection of such distortion in the CMB experiment \cite{Kohri:2014lza}.
Therefore, the primordial perturbations that can explain the abundance of the 
observed SMBHs at high redshifts and, at the same time, produce little CMB spectral distortion 
must be strongly non-Gaussian.

Such a non-Gaussian primordial perturbations may be naturally realized by the quantum fluctuations
of the spectator field created during inflation.
The spectator field is a scalar field which acquires quantum fluctuations during inflation but is not an inflaton and subdominant during inflation.
Approximating the spectator field as a free field, the fluctuations of the spectator field
are Gaussian.
However, thanks to the subdominance of the spectator field, 
it is possible to build a model where
the primordial perturbations generated from the spectator field fluctuations 
become highly non-Gaussian. 
In such a case, production of PBHs occurs at locations corresponding to
the extremely high-sigma peaks of the spectator field fluctuations while suppressing the CMB spectral distortion at a level much below the upper limit set by the
existing CMB experiment.
This production mechanism of PBHs have also been considered and studied in \cite{Nakama:2016kfq, Hasegawa:2017jtk, Hasegawa:2018yuy, Kawasaki:2019iis, Kitajima:2020kig}. 
A characteristic prediction common to all the concrete models studied in these papers 
is the strong clustering of the PBHs.
This feature is due to the fact that the rarer the peaks of the Gaussian fluctuations are, the more such peaks are spatially clustered \cite{Kaiser:1984sw}.

The angular correlation function of PBHs originating from the spectator field fluctuations explained above
has been computed in \cite{Shinohara:2021psq} under the assumptions that (i)~PBHs do not grow substantially by accretion between the time of formation 
and the observation time (i.e. $z\sim 6$)\footnote{
Inclusion of the effect of 
mass growth of PBHs due to accretion amounts to taking smaller value of $M_{\rm PBH}$
in Eq.~(\ref{k-MPBH}). In \cite{Shinohara:2021psq}, it was shown that
any choice of $M_{\rm PBH}$ in the range $10^4 M_\odot \lesssim M_{\rm PBH} \lesssim 10^{10} M_\odot$  
predicts a very large amplitude of the PBH correlation function whose shape depends on its mass and the parameters in the model, but the size does not change qualitatively, and hence the assumption~(i) does not affect our arguments. But we note that the fact that the current samples of SMBHs have been discovered as quasars indicates that they are accreting mass almost at maximum efficiency, at least at the time when they were observed (i.e., $z \sim 6)$.
}
and (ii)~the time evolution of the
PBH correlation function due to mergers and motions of PBHs is negligible
for the scales probed by the observations.
According to \cite{Shinohara:2021psq}, the angular correlation function of PBHs existing between 
the redshift range $z_{\rm low} < z < z_{\rm high}$ is given by
\begin{align}
    \label{07171715}
    w_\mrm{PBH}(\theta) = \int^{R_\mrm{high}}_{R_\mrm{low}} dR_1 \int^{R_\mrm{high}}_{R_\mrm{low}} dR_2\ \frac{3R_1^2}{R^3_\mrm{high}-R^3_\mrm{low}}\,\frac{3R_2^2}{R^3_\mrm{high}-R^3_\mrm{low}}\,\xi_\mrm{PBH}(R_1,R_2,\theta).
\end{align}
Here, $R_\mrm{low}\equiv R(z_\mrm{low})$ and $R_\mrm{high}\equiv R(z_\mrm{high})$ 
are the comoving distance $R(z) \equiv \int_0^z \frac{dz'}{H(z')}$ for $z=z_{\rm low}$ and $z=z_{\rm high}$, respectively, and $\xi_{\rm PBH}$ is the correlation function 
of PBHs given by
\begin{align}
    \label{07171716}
    \xi_\mrm{PBH}(r) \approx \frac{(1+\ep (r))^{\frac{3}{2}}}{\sqrt{1-\ep (r)}}\,\exp\biggl(\frac{\ep (r) \nu^2}{1+\ep (r)}\biggr) - 1.
\end{align}
As this equation shows, the PBH correlation function is determined by two quantities
$\nu$ and $\ep (r)$, which we explain below one by one. 
First, the dimensionless quantity $\nu$ is the measure of rareness of the PBHs:
only locations where the random Gaussian scalar field which is a source of PBHs
takes a value greater than $\nu \sigma$ ($\sigma^2$ is the variance of the scalar field fluctuations) collapse to PBHs after inflation. 
It is related to the abundance of the observed number density of quasars by 
\begin{align}
    \label{07171719}
    \frac{e^{-\nu^2/2}}{\sqrt{2\pi}\,\nu} \sim 4\times 10^{-12} f \biggl(\frac{M_\mrm{PBH}}{10^{10}M_\odot}\biggr)^{3/2}\, \biggl(\frac{n_\mrm{QSO}}{40\,\mrm{Gpc^{-3}}}\biggr)\, ,
\end{align}
where $n_\mrm{QSO}$ is the average (comoving) number density of observed quasars. 
We also consider the possibility that PBHs constitute only a fraction of quasars,
and $f$ represents the fraction of PBHs in quasars, which is defined as $f \equiv n_\mrm{PBH}/n_\mrm{QSO}$ where $n_\mrm{PBH}$ is the average number density of PBHs.
We have taken $n_{\rm QSO}=40~ {\rm Gpc}^{-3}$ as a fiducial value
based on \cite{2018ApJ...869..150M}\footnote{
We have used the luminosity function given by Eq.~(8) in \cite{2018ApJ...869..150M} and integrated it in the magnitude range 
$M_{1450} < -22.25$ assuming all the quasars are at $z=6$.
}.
Notice that the right-hand side of the above equation is extremely small 
even if $f=1$ and, hence, $\nu \gg 1$ always holds. 
This shows that PBHs are sparsely populated at their formation epoch. 

Secondly, the function $\ep (r)$ is the normalized two-point function of the spectator field
fluctuations produced during inflation:
\begin{align}
    \label{eq:cI}
    \ep(r) = \frac{c_I}{k_\mrm{max}^{c_I} - k_\mrm{min}^{c_I}}\,\int^{k_\mrm{max}}_{k_\mrm{min}} \frac{dk}{k}\, k^{c_I}\,\frac{\sin(kr)}{kr}\,,\qquad c_I = \frac{2m^2}{3H_I^2}.
\end{align}
Here $m$ and $H_I$ denote a mass of the spectator field and the Hubble parameter during inflation, respectively.
The minimum comoving wave number, $k_\mrm{min}$, defines the size of the observable region 
in which our perturbation is defined. 
Thus, we take $k_\mrm{min}=H_0$ with $H_0$ being the Hubble parameter at the present epoch. The maximum wave number $k_\mrm{max}$ is related to
the PBH mass by Eq.~(\ref{k-MPBH}) with the left-hand side replaced with $k_{\rm max}$.
$k_{\rm max}$ is determined by concrete inflationary models describing how the scalar field fluctuations are converted into primordial curvature perturbations which source PBHs.
Since we want our present analysis to be general and independent of concrete models, 
we leave $k_{\rm max}$ (or equivalently $M_{\rm PBH}$) as well as
$c_I$ and $f$ as free parameters.

\begin{figure}[t]
        \begin{center}
          \includegraphics[keepaspectratio, scale=0.45]{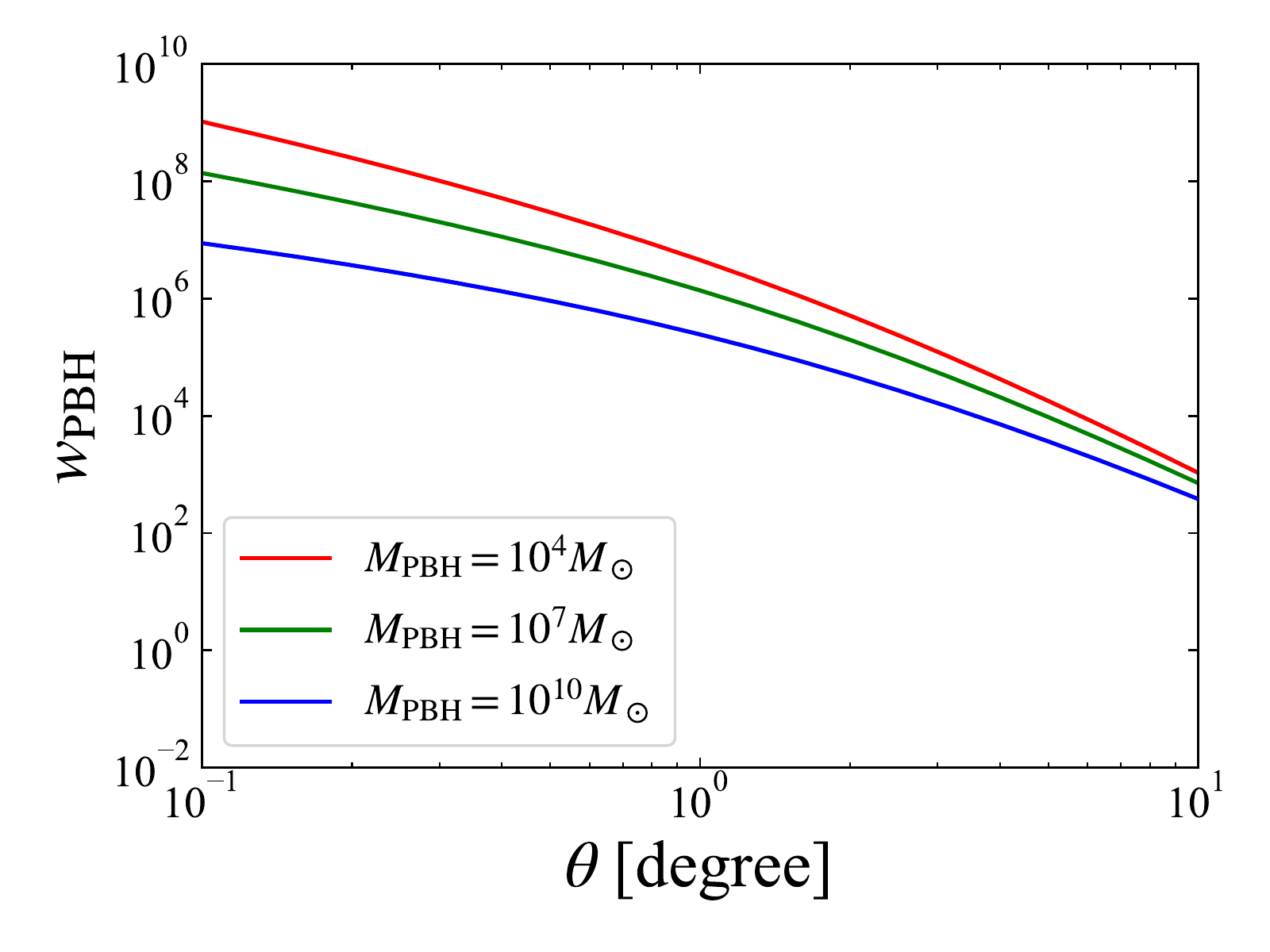}
          \includegraphics[keepaspectratio, scale=0.45]{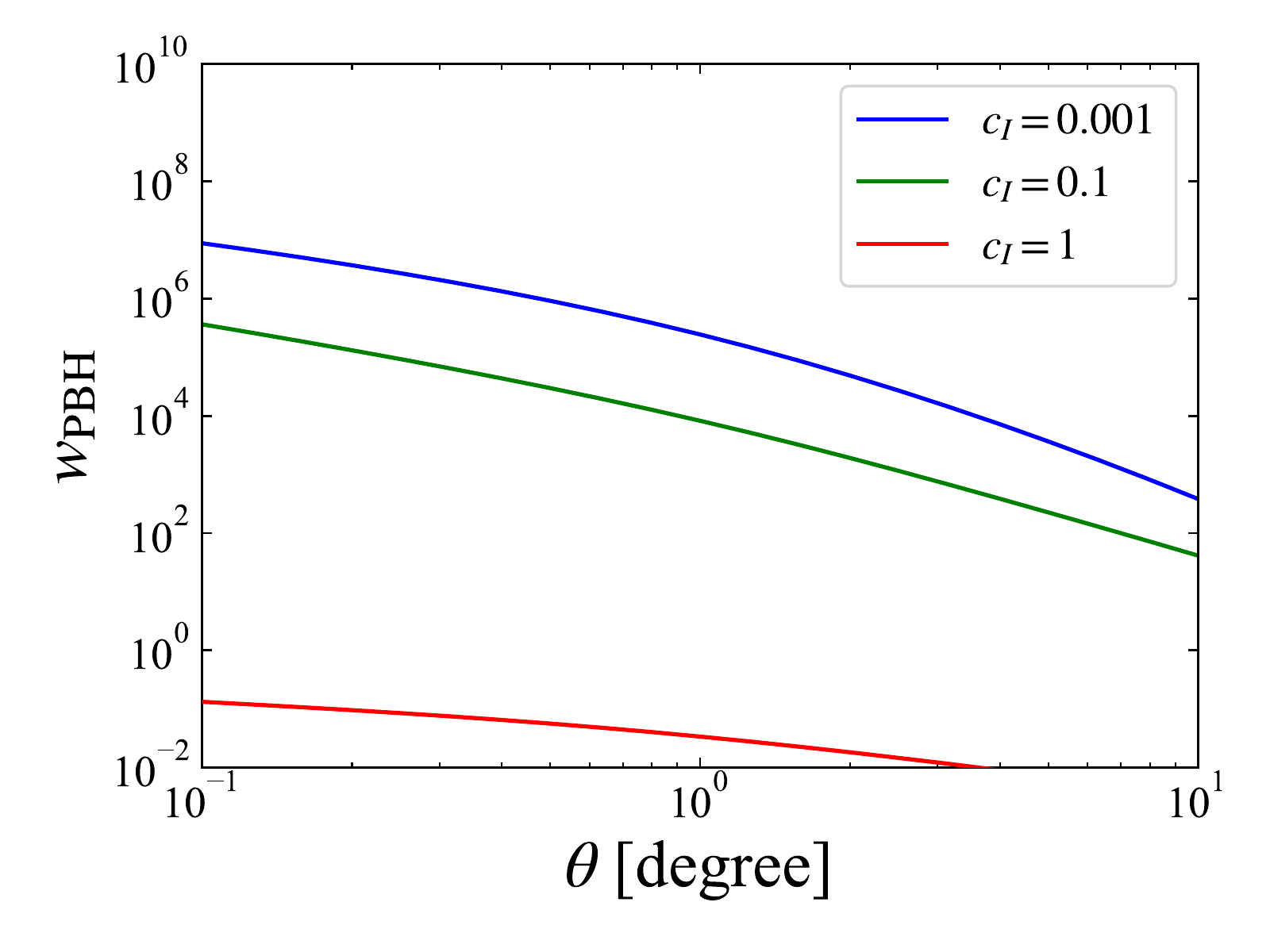}
        \end{center}
      \caption{
      The dependency of the PBH angular correlation function on the PBH mass when $c_I=0.001$ ({\it left panel}) and the scalar field mass when $M_\mrm{PBH}=10^{10}M_\odot$ ({\it right panel}).
      In both cases, the integral to the line-of-sight was performed setting $(z_\mrm{low},\, z_\mrm{high})=(5.88,\,6.5)$.
    \label{fig:w_PBH}
    }
\end{figure}

To summarize, the PBH angular correlation function depends on three quantities 
$(M_{\rm PBH}, c_I, f)$ which are free parameters in our model.
When $f < 1$, the angular correlation function of SMBHs is given by a sum
of the angular correlation function of SMBHs of astrophysical origin
and that of PBHs\footnote{
Actually we need to consider four populations for SMBHs: active PBHs, non-active PBHs, active astrophysical BHs and non-active astrophysical BHs, where ``active" indicates PBHs or BHs with the mass accreting with $M_{1450} < -22~{\rm mag}$. Active populations are expected to be located at more biased places in the dark matter structure where mass accretion is more frequent. However, in the following argument, we assume that the clustering property we measure represents the whole PBHs + astrophysical BHs.
}:
\begin{align}
    \label{eq:omega_th}
    \om = \om_\mrm{ABH} + \om_\mrm{PBH} = (1-f)^2w_\mrm{ABH} + f^2 w_\mrm{PBH}\,,
\end{align}
where $w_\mrm{ABH}$ is the angular correlation function predicted for SMBHs formed by astrophysical processes. In this paper, we compare this angular correlation function with the distribution
of the observed quasars on the sky and derive an allowed region in the parameter space
spanned by $(M_{\rm PBH}, c_I, f)$.
For $\omega_{\rm ABH}$, we do not specify its explicit form in the following analysis since no definite prediction is available. However we note that $\omega_{\rm ABH}$ based on Lyman break galaxies (LBGs) at high redshift has been obtained \cite{Harikane:2015unm}, which can be considered as a lower limit of $\omega_{\rm ABH}$. This is because, since the number of the entire LBG population is dominated by fainter LBGs which have lower stellar mass, the BH mass of LBGs should have systematically lower ones that of SHELLQs quasars.

\section{Quasar sample and clustering analysis \label{sec:analysis}}
\subsection{Sample of $z\sim6$ quasars}

\begin{figure}[ht]
        \begin{center}
          \includegraphics[keepaspectratio, scale=0.6]{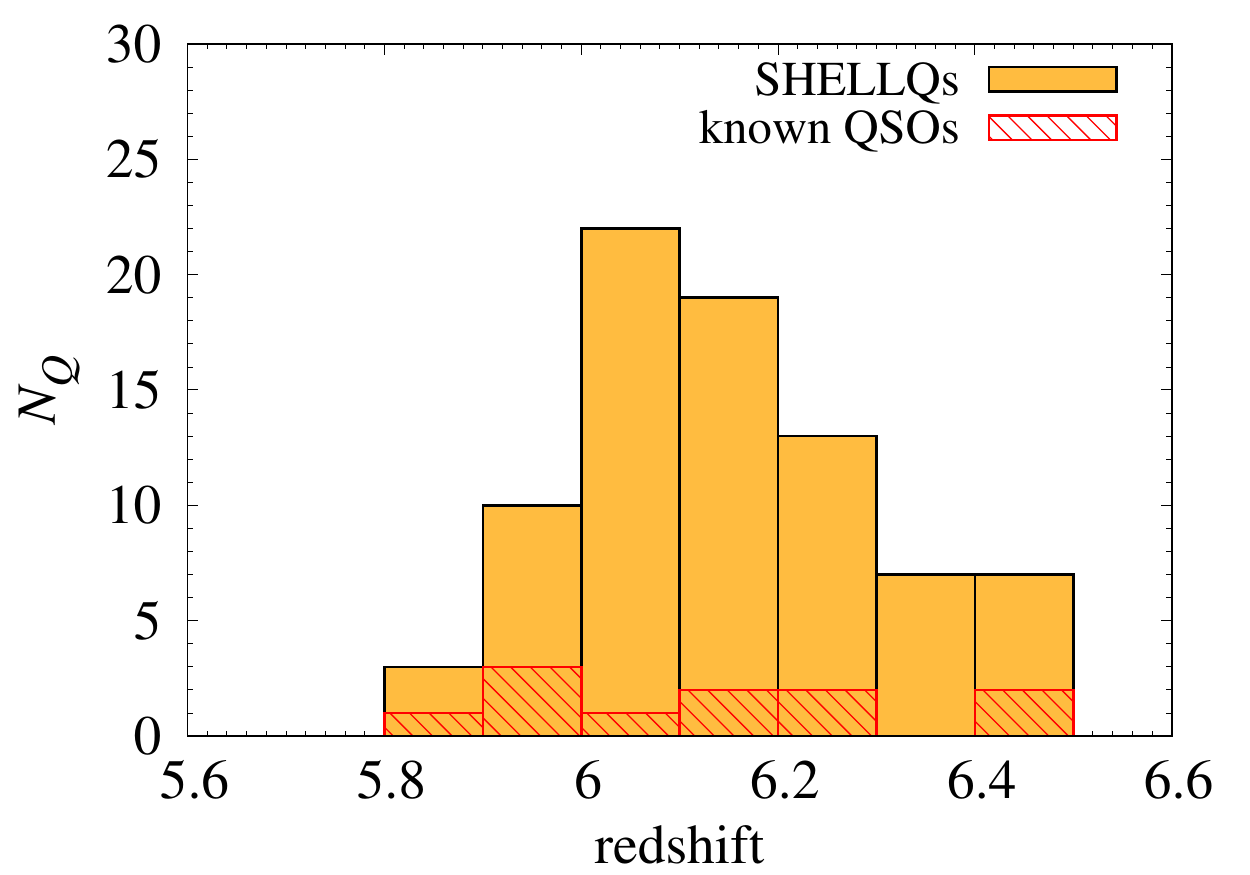}
           \includegraphics[keepaspectratio, scale=0.6]{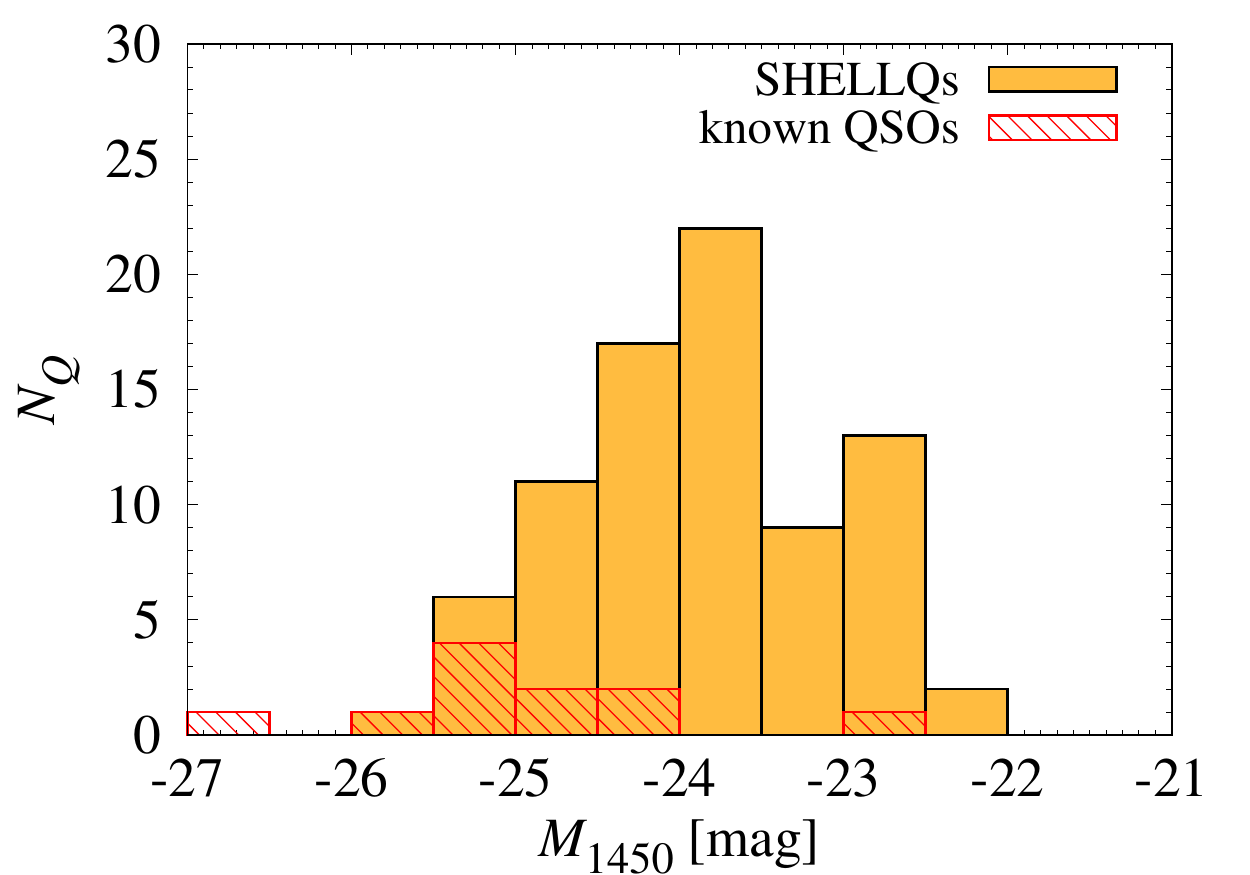}
        \end{center}
      \caption{
      Distributions of redshift (left) and UV absolute magnitude $M_{1450}$ (right) of quasars we adopt in our analysis.
    \label{fig:number_QSOs}
    }
\end{figure}

In order to evaluate the ACF of the quasars at $z\sim6$ with the projected two-point angular correlation function, we use the quasar sample constructed by Matsuoka et al. (SHELLQs)~\cite{Matsuoka:2021jlr}.
It contains 81 quasars covering a wide magnitude range of $-25.58\leq M_{1450}\leq-22.25$ at $5.88 \leq z \leq 6.49$, selected from the HSC-SSP PDR3 Wide-layer dataset with a sky coverage of $\sim1200\,\mrm{deg}^2$ \cite{Aihara:2021jwb}.
Additionally, to increase sample size, we include 11 known quasars \cite{Willott:2013gyg,willott2009six,jiang2009survey,Jiang:2007fz,mazzucchelli2017physical,Venemans:2015rca,2016ApJS..227...11B,kim2015discovery,kashikawa2014subaru,zeimann2011discovery} falling within the sky coverage of the Wide-layer dataset in the same redshift range.
Distributions of redshift and UV absolute magnitude of the quasar sample are plotted in the left and right panel of Fig.~\ref{fig:number_QSOs}, respectively.

The two-point correlation function is estimated by comparing the number of pairs of real quasars and that of randomly-distributed mock quasars in the survey area.
We construct the sample of random quasars as follows. 
HSC-SSP PDR3 provides a catalog of random points with a density of 100 points per $\mrm{arcmin}^2$ for each coadd image for each filter \cite{Aihara:2021jwb,Aihara:2019xyr}.
To mask regions with unreliable photometry, we apply the same criteria used to select the real quasars to the random catalog:
\begin{align*}
\mrm{pixelflags\_edge} &= \mrm{False} \\
\mrm{pixelflags\_saturatedcenter} &= \mrm{False} \\
\mrm{pixelflags\_crcenter} &= \mrm{False} \\
\mrm{pixelflags\_bad} &= \mrm{False}\\
\mrm{isprimary} &= \mrm{True}\,,
\end{align*}
in the $izy$-bands, and
\begin{align*}
\mrm{inputcount\_value} \geq 2
\end{align*}
in the $z_\mrm{AB}$ band.
In addition, we apply the following criteria to mask regions around bright sources,
\begin{align*}
\mrm{mask\_brightstar\_halo} &= \mrm{False}  \\
\mrm{mask\_brightstar\_ghost} &= \mrm{False} \\
\mrm{mask\_brightstar\_blooming} &= \mrm{False}\,,
\end{align*}
in the $izy$-bands, and
\begin{align*}
\mrm{mask\_brightstar\_channel\_stop} &= \mrm{False} 
\end{align*}
in the $y_\mrm{AB}$ band.
Then we randomly select 100,000 points.

\begin{figure}[ht]
\begin{center}
    \includegraphics[clip,width=12.0cm]{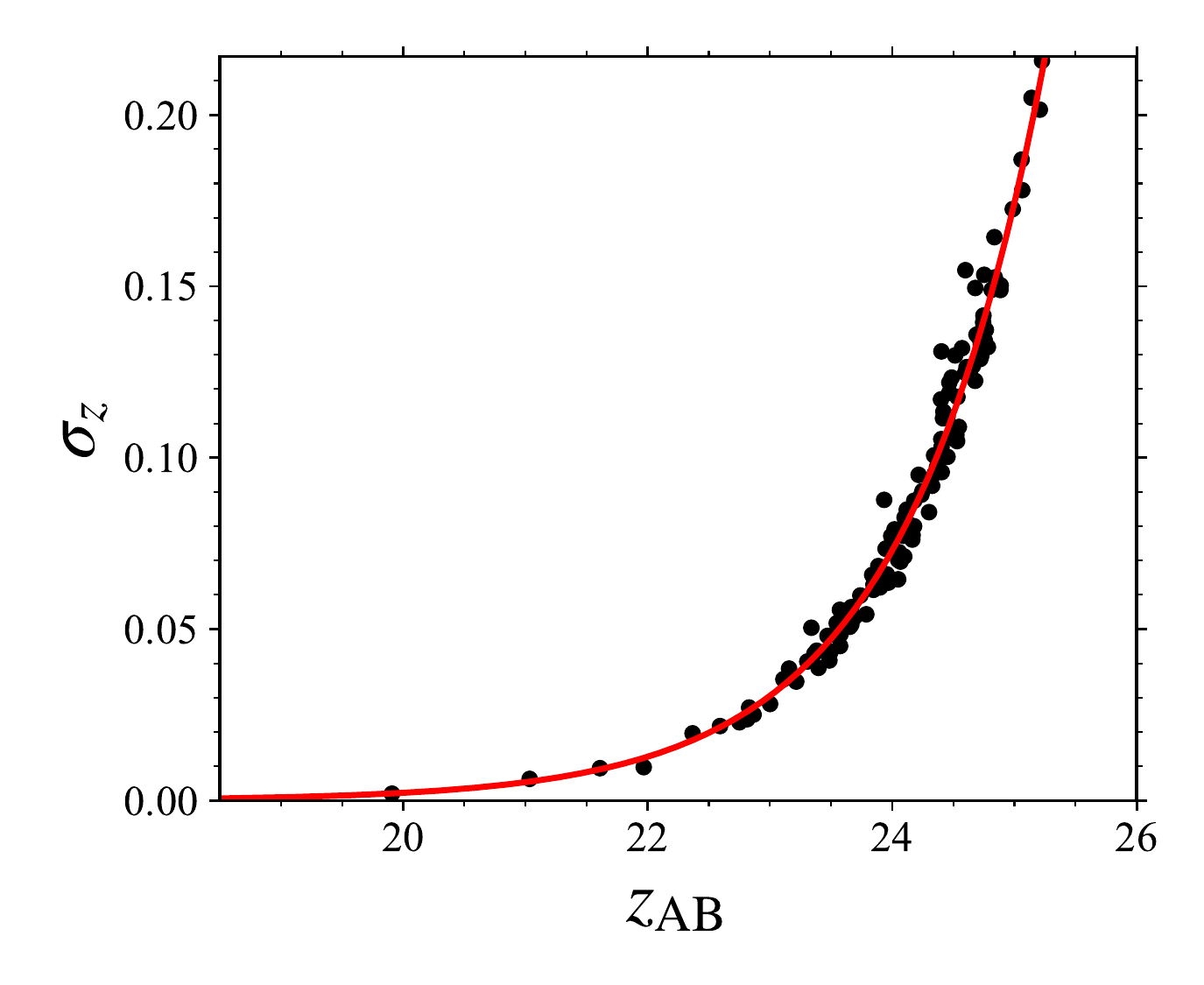}
    \caption{
    Relations between the $z$-band photometric uncertainty and $z$-band magnitude. Dots represent point sources on Patch (2,3) of Tract 9567 of HSC PDR3 Wide imaging. Red line shows the fitting formula, $\si_z=a\exp(bz_\mrm{AB})$ where $a=(6.03\pm2.26)\times10^{-11}$ and $b=0.87\pm0.02$.
\label{fig:mag-error}
}
  \end{center}
\end{figure}

We note that the detection limit does not uniformly distribute over the sky coverage of the Wide-layer dataset, as PDR3 only includes data obtained by January 2020, and the detection limit can be affected by non-uniform seeing. 
Since the non-uniform distribution of detection limits can affect the selection of real quasars, it is necessary to reproduce this non-uniform distribution in the random quasars as well. 
We consider this effect as follows. Firstly, we examine the relation between magnitude errors and magnitudes in the $z$-band of real point sources on the Wide-layer dataset. 
We adopt the criteria used in \cite{2018ApJ...869..150M}, $z_\mrm{psf,AB}-z_\mrm{CModel,AB}<0.15$ where $z_\mrm{CModel,AB}$ is the CModel magnitude \cite{bosch2018hyper} to select point sources, and found that Patch (2,3) of Tract 9567 includes 126 point sources, which are abundant enough for the analysis\footnote{
Tract is a $1.7\times1.7$ degree region, and each tract is divided into $9\times9$ patches.
}.
An exponential function is applied to fit magnitude errors in the $z$-band of the real point sources on the patch as a function of their $z$-band magnitudes. 
The best-fit model is plotted by red line in Fig.~\ref{fig:mag-error}. 
Then, we shift the best-fit model to match the 5-sigma detection limit in the $z$-band for each patch, and assign the corresponding magnitude error at $z_\mrm{AB}=24.5$ to each random position falling in the patch. 
Finally, we adopt the criterion in the $z$-band magnitude error $\sigma_z<0.155$, which is used to select the real quasars, to the random positions criteria \cite{Matsuoka:2021jlr}.

\begin{figure}[ht]
\begin{center}
    \includegraphics[clip,width=12.0cm]{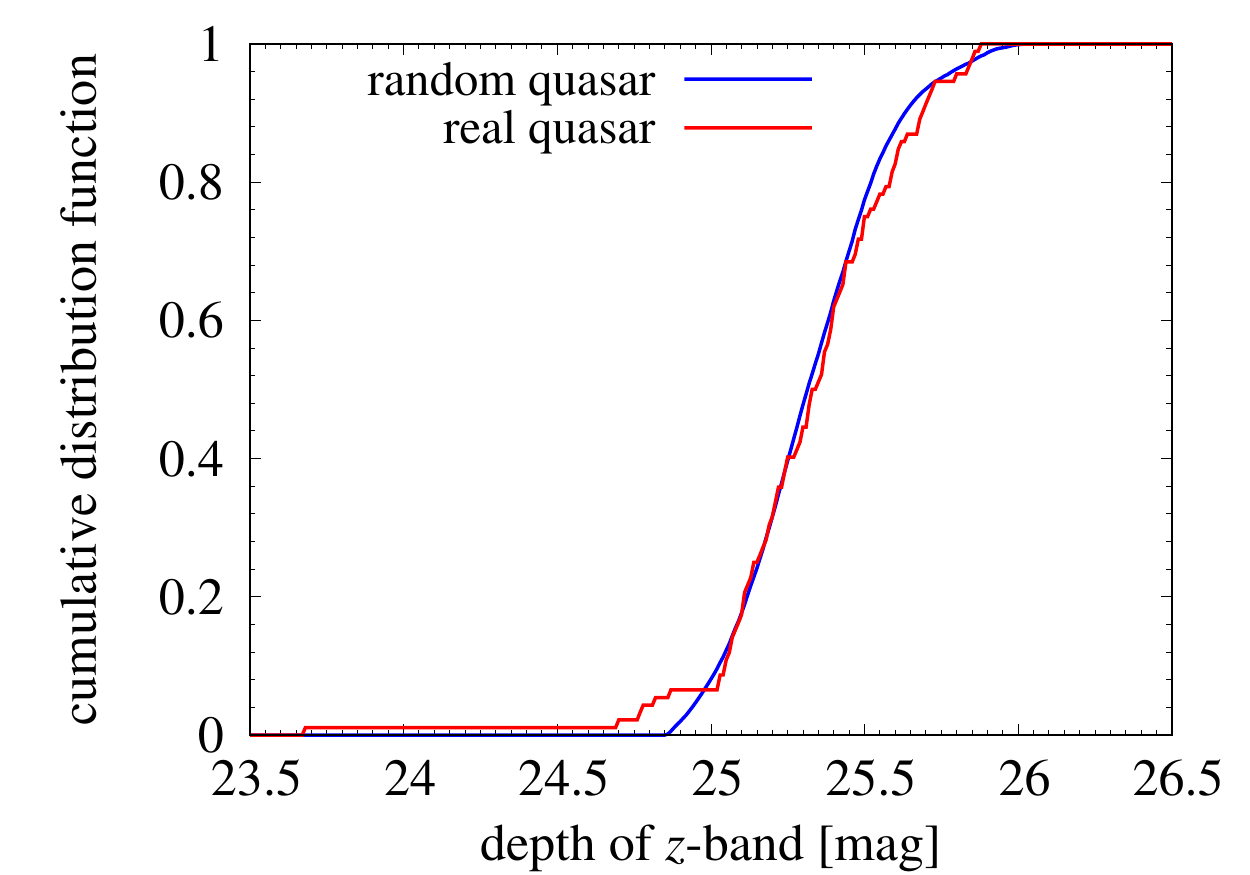}
    \caption{
    Cumulative distributions of the $z$-band detection depth at the patch of the real (red) and random (blue) quasars.
\label{fig:cdf}
}
  \end{center}
\end{figure}

In total, we obtain 87,933 random positions, and we regard them as random quasars hereafter. 
Fig.~\ref{fig:cdf} shows the cumulative distribution functions of the depth in the $z$-band for random and real quasars. 
Although there are small discrepancies at the depth around $24.8$ and $25.6$ mag, given the fraction of affected patches are limited, both distributions are considered to be consistent, i.e., the random quasars are properly generated.
We confirmed that the size of random quasars is sufficient for the clustering analysis as more random quasars do not significantly change the derived angular correlation function.

\subsection{Angular correlation function of $z\sim6$ quasars}

To derive the angular correlation function $\omega$ with $z\sim6$ quasars, we adopt the estimator given by Davis~\&~Peebles~\cite{Davis:1982gc}:
\begin{align}
\label{eq:omega_ob}
\omega(\theta) = \frac{DD(\theta)}{DR(\theta)} - 1\,,
\end{align}
where $DD$ and $DR$ are the normalized counts of the quasar-quasar and quasar-random pairs between $\theta - \De \theta$ and $\theta + \De \theta$, and they are given by
\begin{align}
\label{eq:DD_and_DR}
DD(\theta) = \frac{N_{QQ}(\theta)}{N_Q(N_Q-1)/2}\,, \qquad DR(\theta) = \frac{N_{QR}(\theta)}{N_QN_R}\,.
\end{align}
Here, $N_{QQ}$ and $N_{QR}$ are the numbers of the quasar-quasar and quasar-random pairs in the each bin, and $N_Q$ and $N_R$ are the total counts of the real and random quasars, respectively.
We count the number of $DD$ or $DR$ pairs in 12 logarithmically-spaced bins in the angular scale of $0.2<\theta/\mrm{deg}<10$.

\begin{figure}[ht]
\begin{center}
    \includegraphics[clip,width=11.0cm]{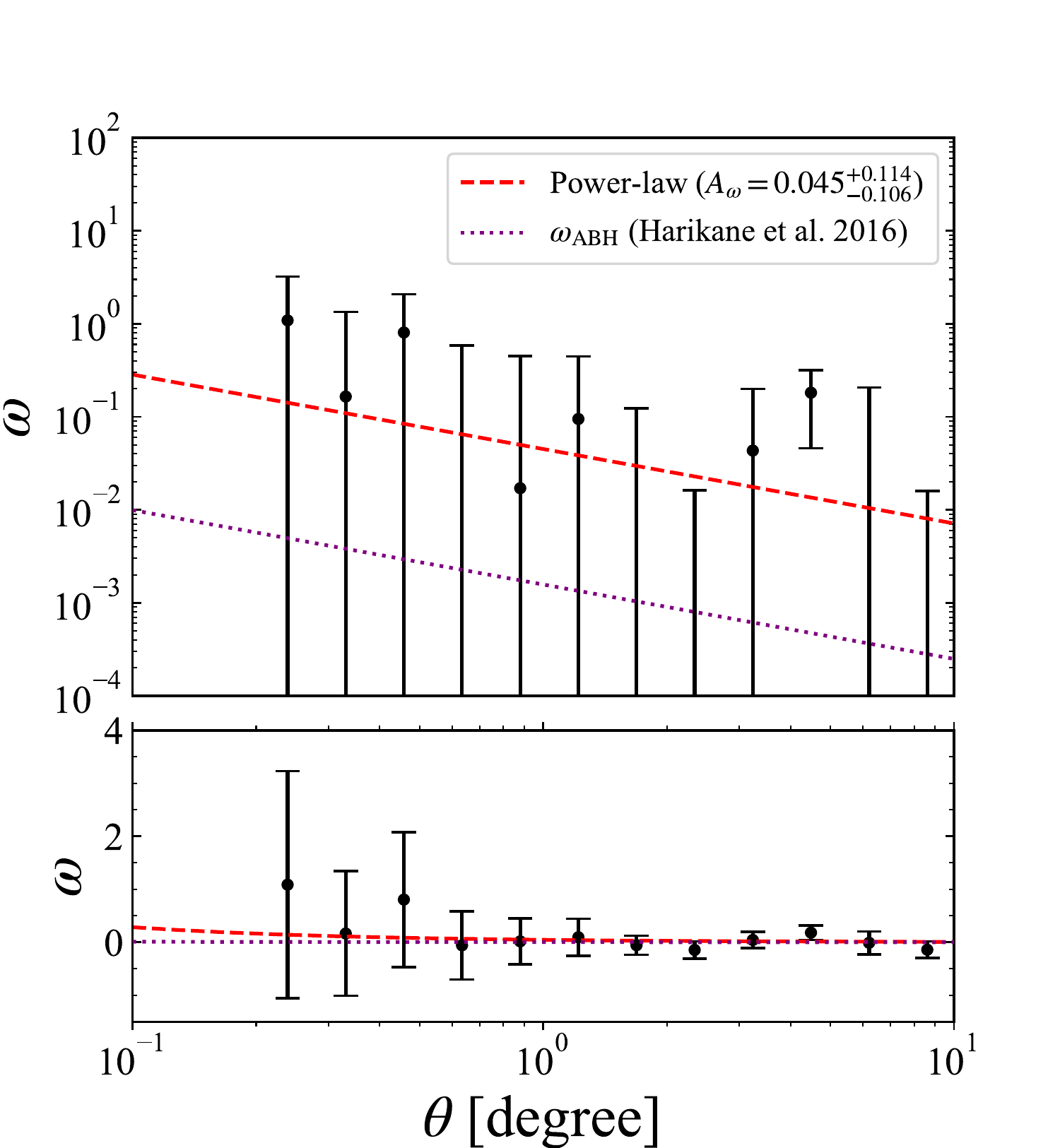}
    \caption{
    Two-point angular auto-correlation function derived with $z\sim6$ quasars (black dot) in the logarithmic (upper panel) and linear (bottom panel) scales. The best-fit power-law model is shown by red dashed line. For comparison, we also display the ACF of LBGs down to $m_\mrm{UV}=28.4$ at $z\sim6$ \cite{Harikane:2015unm} by purple dotted line.
\label{fig:2pcf}
}
  \end{center}
\end{figure}

We evaluate the uncertainty of ACF by the Jackknife resampling \cite{SDSS:2004oes}. 
We separate the entire survey area into 10 sub-regions. 
In each resampling, we exclude one sub-region, and estimate the ACF for the real and random quasars on the remaining regions with the same estimator described above. 
The uncertainty of the ACF is evaluated by the diagonal elements of the covariance matrix:
\begin{align}
C_{ij} = \frac{N-1}{N}\, \sum^{N}_{k=1} \bigl(\om^k_i - \overline{\om_i}\bigr) \bigl(\om^k_j - \overline{\om_j}\bigr)\,,
\end{align}
where $\overline{\om_i}$ is the average of the angular correlation function for 10 estimates. We regard the square root of the diagonal elements $C_{ii}$ as the uncertainty for the $i$-th bin, i.e., $\si_{\mrm{JK},i}\equiv \sqrt{C_{ii}}\,$.
The resulting ACF and its uncertainty for each bin are plotted in Fig.~\ref{fig:2pcf}, and summarized in Table~\ref{table:summary}. 
\begin{table}[ht]
  \caption{ACF of $z\sim6$ quasars}
  \label{table:summary}
  \centering
  \begin{tabular}{|c|ccrrcc|}
    \hline
    \rule{0cm}{0.5cm}No. & $\theta$\,[deg] & $(\theta-\De\theta,\, \theta+\Delta\theta)$\,[deg]  & $N_{QQ}$  &  $N_{QR}$ & $\om$ & $\si_\mrm{JK}$  \\[2pt]
     \hline \hline
    \rule{0cm}{0.5cm}
    1 & 0.24 & $(0.2 ,\,0.28)$  & 1\ & 926 & 1.087 & 2.143\\[2pt]
    2 & 0.33 & $(0.28,\,0.38)$  & 1 & 1659 & 0.165 & 1.178\\[2pt]
    3 & 0.46 & $(0.38,\,0.53)$  & 3 & 3211 & 0.806 & 1.273\\[2pt]
    4 & 0.63 & $(0.53,\,0.74)$  & 3 & 6151 & -0.057 & 0.643\\[2pt]
    5 & 0.88 & $(0.74,\,1.02)$  & 6 & 11401 & 0.017 & 0.434\\[2pt]
    6 & 1.22 & $(1.02,\,1.41)$  & 12 & 21185 & 0.095 & 0.350\\[2pt]
    7 & 1.69 & $(1.41,\,1.96)$  & 19 & 38889 & -0.056 & 0.179\\[2pt]
    8 & 2.34 & $(1.96,\,2.71)$  & 30 & 67589 & -0.146 & 0.162\\[2pt]
    9 & 3.24 & $(2.71,\,3.76)$  & 59 & 109289 & 0.043 & 0.155\\[2pt]
    10 & 4.49 & $(3.76,\,5.21)$ & 96 & 157019 & 0.181 & 0.136\\[2pt]
    11 & 6.21 & $(5.21,\,7.22)$ & 104 & 203253 & -0.011 & 0.217\\[2pt]
    12 & 8.61 & $(7.22,\,10.0)$ & 108 & 242998 & -0.141 & 0.157\\[2pt]
    \hline
  \end{tabular}
\end{table}

We fit the derived ACF with power-law function:
\begin{align}
    \label{eq:power-law}
    \omega_\mathrm{fit}(\theta) = A_\omega \biggl(\frac{\theta}{\mrm{deg}}\biggr)^{-\beta}\,.
\end{align}
Due to the limited size of quasar samples used in the analysis, we fix the power index $\beta$ to 0.8 to reduce the number of free parameters in the fitting. 
The chosen power index is determined by the ACFs of galaxies in the lower redshifts \cite{ouchi2010statistics,Harikane:2015unm,Ouchi:2001gw,Ouchi:2003xw,Foucaud:2003xd,Foucaud:2010hu}.
We carry out the fitting through the maximum likelihood (ML) method \cite{1997MNRAS.291..305C} (see also \cite{He:2017crq}), which is less affected by negative bins compared to the $\chi^2$ fitting method, since the former method does not require a specific binning.

We define the likelihood $\mcal{L}$ of having the observed pair sample against the model prediction under the assumption that the pair count in each bin follows the Poisson distribution as follows \cite{1997MNRAS.291..305C},
\begin{align}
\mcal{L} \equiv \prod^{\mrm{all\ bins}}_{i=1} \frac{e^{-h(\theta_i)}\bigl[h(\theta_i)\bigr]^{N_{QQ}(\theta_i)}}{\bigl[N_{QQ}(\theta_i)\bigr]!}\,,
\end{align}
where $h_i\equiv\bigl(1+w_i\bigr)DR_i$\, is the expected quasar-quasar pair count estimated by the quasar-random pair counts within the small interval around $\theta$. 
The free parameter $A_\om$ is determined by minimizing $S \equiv -2\ln \mcal{L}$ which is explicitly given by 
\begin{align}\label{eq:ML}
S = 2\sum^{\mrm{all\ bins}}_i h(\theta_i) - 2\sum^{\mrm{all\ bins}}_i N_{QQ}(\theta_i) \ln h(\theta_i)\,.
\end{align}
If we assume $S$ follows the $\chi^2$ distribution, the 1$\sigma$ uncertainty of the best-fit $A_\om$ can be evaluated by the range with $\Delta S=S-S_\mrm{min}$ less than one. 
The resulting best-fit $A_\om$ is $0.045^{+0.114}_{-0.106}$, and the best-fit power law model is plotted by red dashed line in Fig.~\ref{fig:2pcf}. 
Arita~et~al.~\cite{Arita2023} have also studied the ACF of quasars at $z\sim6$ using data from the SHELLQs project\footnote{
Arita et al.~\cite{Arita2023} performed the analysis independently from this work.
}. Here we comment on the comparison between that derived by Arita et al. and ours. Their quasar sample excluded type-2 AGN candidates and the final sample consists of 107 quasars with a luminosity range of $-27.0<M_{\rm 1450}<-21.3$. While there were slight differences between their sample and ours, and they adopted a different estimator from Landy \& Szalay \cite{Landy:1993yu} to calculate the ACF, we found that the best-fit power-law model of the ACF of quasars at $z\sim6$ is consistent between the two studies.

\section{Testing PBH scenario as SMBHs\label{sec:constraint}}

Now we discuss constraints on the PBH scenario as SMBHs using the angular correlation function obtained from SHELLQs quasar data. 
\begin{figure}[H]
         \begin{center}
          \includegraphics[keepaspectratio, scale=0.6]{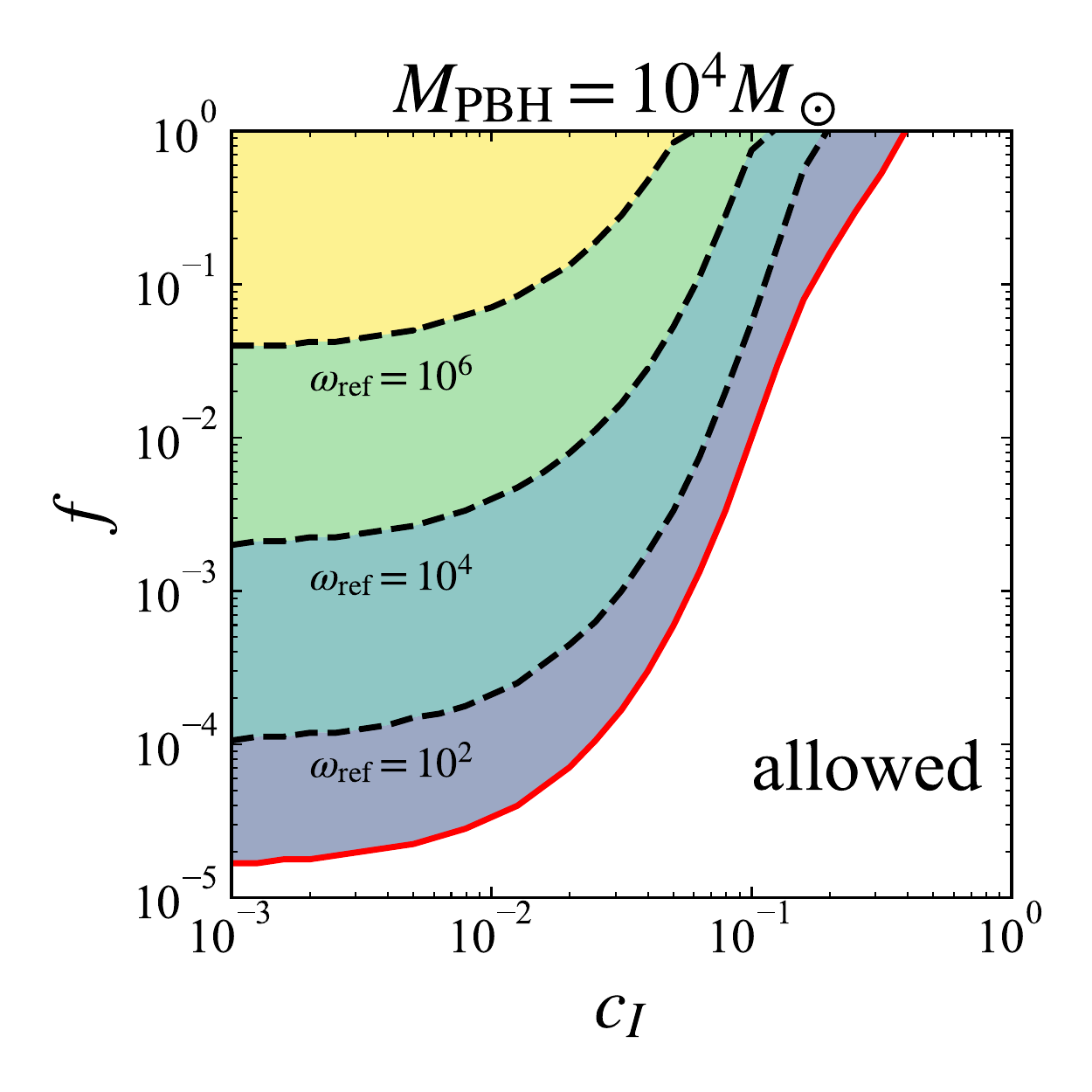}
          \includegraphics[keepaspectratio, scale=0.6]{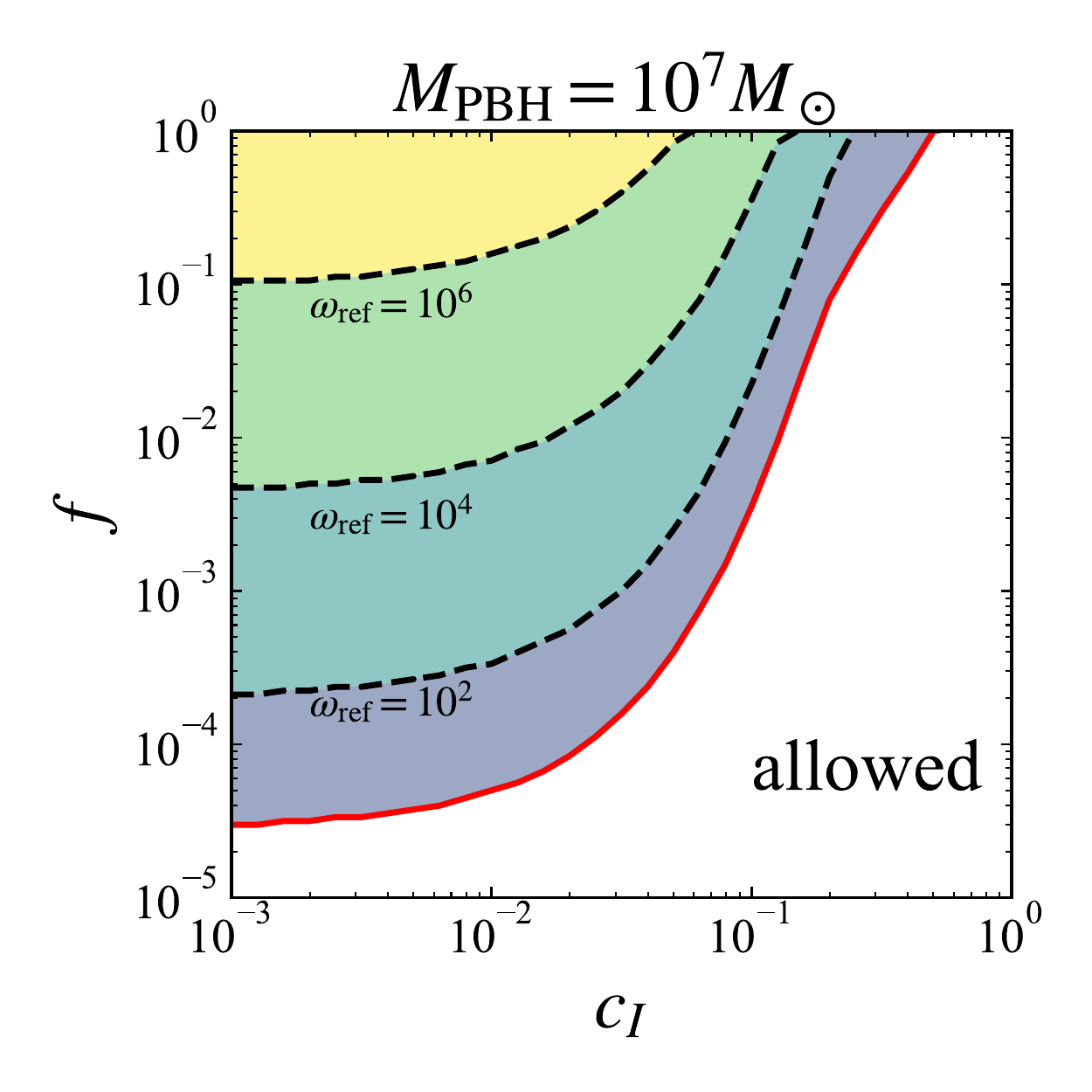}\\
          \includegraphics[keepaspectratio, scale=0.6]{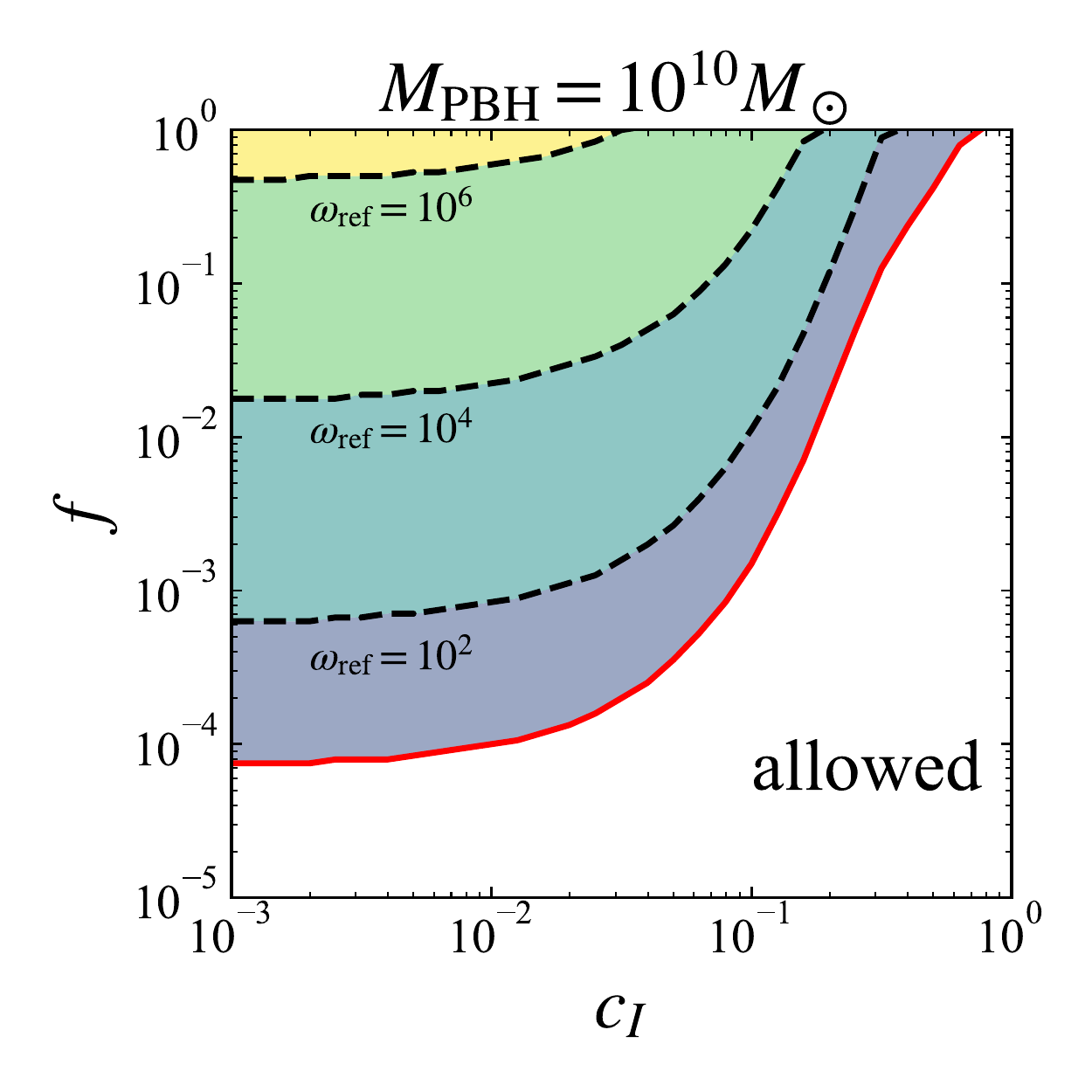}
           \end{center}
      \caption{
      Constraint from the angular correlation function derived from SHELLQs quasar data on the $c_I$-$f$ plane. White region is allowed from the requirement that $\om_{\rm PBH} < \om_\mrm{obs}$ at $2\si$ in all bins. Cases with $M_\mrm{PBH}=10^4 M_\odot$ (top left panel), $10^7 M_\odot$ (top right panel) and $10^{10} M_\odot$ (bottom panel) are shown. The dashed lines show the value of $\om_{\rm ref}$ predicted by the PBH model at the minimum angular scale among 12 bins, i.e., $\om_\mrm{ref}\equiv\om(0.24^\circ)$.
    \label{fig:2pcf_constrain}
    }
\end{figure}
Fig.~\ref{fig:2pcf_constrain} shows 2$\sigma$ constraints on the PBH parameters $c_I$ and $f$ for different values of $M_{\rm PBH}=10^4 M_\odot$, $10^7 M_\odot$ and $ 10^{10}M_\odot$. We judge that a model is allowed when it predicts the ACF consistent with the observed ACF within 2$\sigma$ error for all bins.
White region in each panel is the allowed one consistent with the SHELLQs data and
the left region from red curve  (shaded with colors) is ruled out. To obtain the constraints,  we neglect the contribution from $\omega_{\rm ABH}$, which gives a conservative limit on the PBH parameters. We also show contours of the value of $\om_{\rm ref}$ predicted by the PBH model at the minimum angular scale among 12 bins, i.e.,  $\om_\mrm{ref}\equiv\om(0.24^\circ)$.

From this result, we can emphasize two important points: 
firstly, three panels look similar, which means that our constraint is only weakly
dependent on $M_{\rm PBH}$. 
This can be understood from the left panel of Fig.~\ref{fig:w_PBH} which shows that,
compared to the right panel of Fig.~\ref{fig:w_PBH},
the PBH angular correlation function does not change significantly as we vary
$M_{\rm PBH}$ by several orders of magnitude.
Although it is still uncertain that how much the PBHs can grow their
masses through accretion which may occur until PBHs are formed at high redshifts,
our constraint on the PBH scenario as the origin of the SMBHs at high redshifts
is rather insensitive to such an uncertainty.
Secondly, the possibility that PBHs comprise all the observed SMBHs is completely ruled out
except the case where $c_I$ is close to $1$.
Especially, the contribution of PBHs is severely restricted to be 
$f \lesssim 10^{-4}$ for the case where
the scalar field is nearly massless ($c_I \ll 1$).
This is consistent with a crude estimate obtained by requiring that 
the angular correlation function of PBHs at the observationally relevant angular scales 
$\omega_{\rm PBH} \sim 10^8$ (see Fig.~\ref{fig:w_PBH}) 
should not exceed the observed values $\omega_{\rm obs} \sim 0.1$
(see Fig.~\ref{fig:2pcf}): 
$f \lesssim \sqrt{\frac{\omega_{\rm obs}}{\omega_{\rm PBH}}}={\cal O}(10^{-4})$.

Our result suggests that the PBH models to explain the SMBHs at high redshifts
from non-Gaussian fluctuations generated by Gaussian fluctuations of nearly massless spectator
field are incompatible with the observed spatial distribution of the SMBHs.
Since this conclusion essentially comes from the property
of the spectator field fluctuations that the locations of rare peaks of 
the field fluctuations, which are nothing but
the locations of PBHs, are strongly clustered, our constraint does not rely on
the concrete mechanism to convert the spectator field fluctuations
to the primordial curvature perturbations.
Thus, our conclusion can be applied to any PBH models as long as Gaussian fluctuations of a nearly massless spectator field give the seeds of PBH formation.

On the other hand, the white region in each panel of Fig.~\ref{fig:2pcf_constrain} is currently 
consistent with observations. A PBH scenario is still viable 
if the mass of the spectator field
is comparable to the Hubble parameter during inflation.  The constraint would get severer by reducing the error bars of the measurements of the angular correlation function of high-$z$ SMBHs.
This will be achieved in future by detecting more number of quasars. 
We will discover more than 200 quasars at $z > 6$ in total when the SHELLQs is completed. A significantly larger number of high-$z$ quasars would be discovered in future projects, such as Rubin LSST \cite{LSST}.
It should also be noted that our analysis of SHELLQs data provides a hint 
that the observed angular correlation function is non-zero.
If the angular correlation function is actually non-zero, the models predicting the 
Poisson distribution of PBHs are ruled out. 
For instance, the PBHs originating from the vacuum bubbles produced by
quantum tunneling during inflation \cite{Deng:2017uwc} are not compatible with observations
if the tunneling rate depends only on the adiabatic direction (see footnote~\ref{f-bub}).
These considerations suggest that further reduction of the measurement errors 
of the angular correlation function by using more number of quasars in future
is crucial to test the possibility of PBHs as the origin of the observed SMBHs at high redshifts.

\begin{figure}[ht]
\begin{center}
    \includegraphics[clip,width=12.0cm]{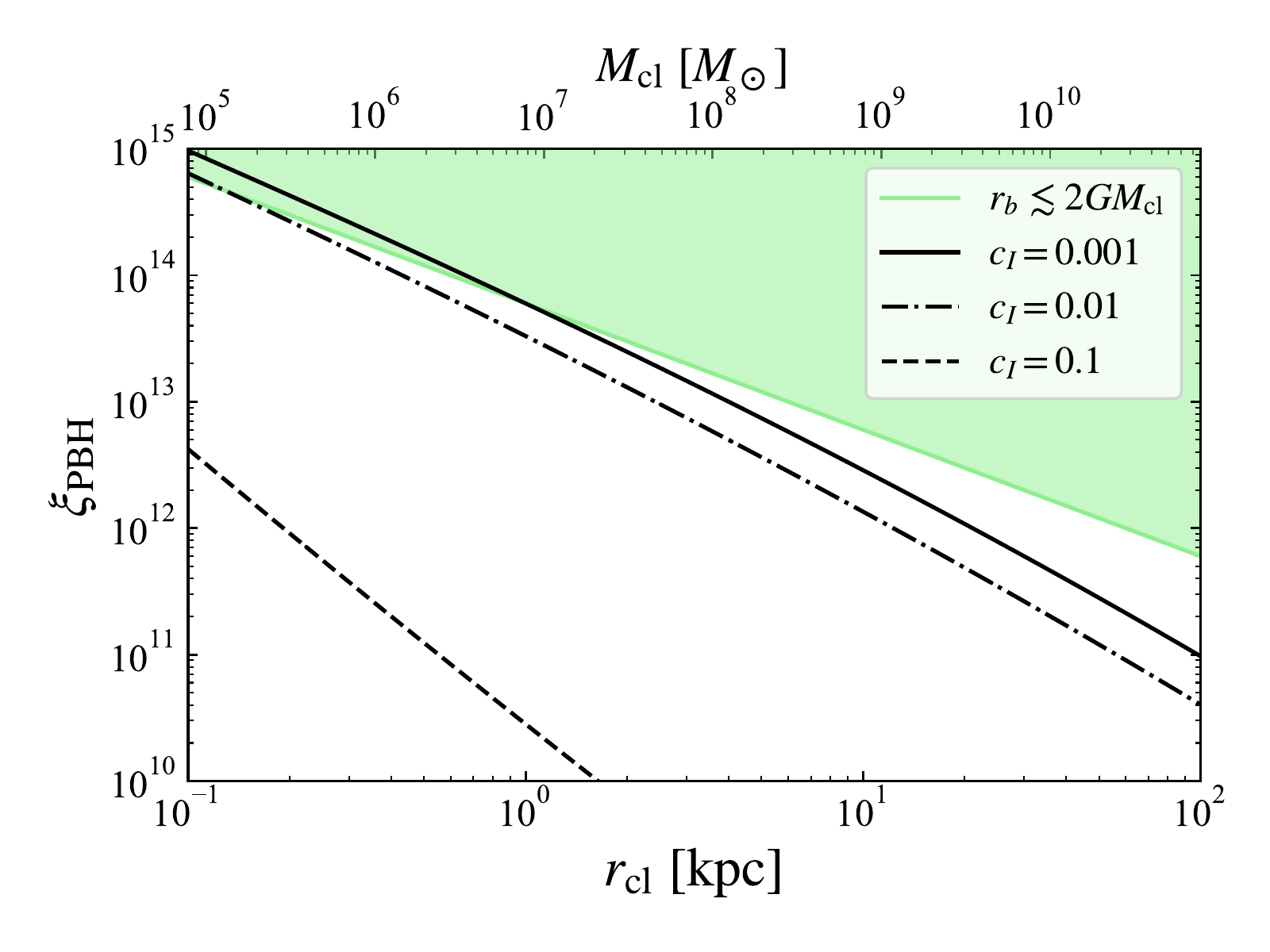}
    \caption{
    The PBH correlation function in the scenario that PBH clump collapses into SMBH. The green line is the lower bound given from the condition that the clump radius $r_b$ is shorter than the Schwarzschild radius.
\label{fig:direct-collapse}
}
\end{center}
\end{figure}

Before closing this section, let us mention that our result also severely constrains
our PBH model using the spectator field in the context of
an alternative scenario \cite{DeLuca:2022bjs} in which SMBHs at high redshifts originate 
from the direct collapse of PBH clumps containing many smaller PBHs. 
In this scenario, the initial PBH number density inside a clump is so large that 
the clump directly turns into a much heavier BH 
soon after the clump decouples 
from the Hubble expansion in deep radiation dominated era as is the case in 
the standard PBH formation in which overdense region with large amplitude decouples 
from the Hubble expansion soon after the horizon reentry and collapses to a BH. 
At the level of the order-of-magnitude estimate, 
the direct collapse of the PBH clump is achieved if the physical size of the clump 
at the time of the decoupling is less than the Schwarzshild radius of the clump,
which we use as the criterion for the direct collapse. 
As a concrete example, green line in Fig.~\ref{fig:direct-collapse} 
represents the boundary of the criterion in the case where $M_{\rm PBH}=M_\odot$
and $f_{\rm PBH}=10^{-9}$\,\,\footnote{
In \cite{DeLuca:2022bjs}, the redshift when
the decoupling of the clump from the Hubble expansion occurs is fixed to 
$z_{\rm eq}/\xi_{\rm PBH}$, where $z_{\rm eq}$ is the redshift at the
matter-radiation equality. 
On the other hand, the green line in Fig.~\ref{fig:direct-collapse}
is obtained by adopting $z_{\rm eq}/(\xi_{\rm PBH}f_{\rm PBH})$ as the redshift of decoupling.
}. 
The direct collapse occurs in the region above the green line. 
The horizontal and vertical axes are the comoving radius of the clumps and 
the magnitude of the two-point correlation function, respectively. 
Upper horizontal axis represents the masses of the resultant heavier BHs originating 
from clumps. 
As we can verify, $\xi_{\rm PBH}$ intersects with the green line 
at $r_{\rm cl} \sim 1~{\rm kpc}$ if $c_I$ is 
sufficiently small ($c_I \lesssim 0.01$).
The mass of the resultant heavier BHs is $\simeq 10^7~M_\odot$. 
This magnitude is large enough to explain the origin of the SMBHs 
observed at high redshifts without resorting to the efficient mass growth due
to accretion after the formation of the heavier BHs.
However, as we see in the figure,
in such cases, the magnitude of the correlation function 
at length scales corresponding to the quasar measurements largely exceeds the measured 
correlation function obtained in this paper from the quasar measurements.
Thus, the scenario proposed in \cite{DeLuca:2022bjs} in the context of our PBH model
is incompatible with the measured quasar distribution on the sky
and cannot be considered to be viable.

\section{Conclusion  \label{sec:conclusion}}

We derived the angular correlation function by using 92 quasars at  $z\sim6$ observed by SHELLQs, and the result is shown in Fig.~\ref{fig:2pcf}. We fitted the derived ACF to the power-law function~\eqref{eq:power-law} and obtained its best-fit model as $A_\omega = 0.045^{+0.114}_{-0.106}$. Then we compared the angular correlation functions from SHELLQs with the theoretically predicted ones from the PBH model, which can serve as a critical test for the PBH SMBH scenarios. 

Theoretical predictions for the angular correlation function of PBHs have been given in \cite{Shinohara:2021psq} for a general scenario where Gaussian fluctuations of the spectator field result in highly non-Gaussian curvature perturbation and the perturbation collapses into PBH.
Such a PBH scenario can explain SMBHs while avoiding the constraint from the CMB $\mu$-distortion.
According to \cite{Shinohara:2021psq}, very high clustering of PBHs occurs as long as the mass of the spectator field is sufficiently small. 
However, when the mass gets close to the Hubble parameter during inflation, the size of the angular correlation function becomes smaller (see the right panel of Fig.~\ref{fig:w_PBH}).
In addition, given that observed quasars may be a combination of PBH and astrophysical ones, we introduced the fraction of PBH-origin SMBHs $f$ as an extra parameter. Then we placed a limit on these  parameters 
$c_I$ and $f$ by comparing to the correlation function derived from $z\sim6$ quasars observed by SHELLQs \cite{Matsuoka:2021jlr}.

In Fig.~\ref{fig:2pcf_constrain}, the constraint in the $c_I$--$f$ plane is depicted for several values of PBH masses. From the figure, we can see that the result is almost independent of the PBH mass $M_\mrm{PBH}$. 
Our result indicates that the PBH scenario as SMBHs is excluded 
when the mass of the spectator field is sufficiently smaller than the Hubble rate during inflation.
We note that the astrophysical contribution $\omega_\mrm{ABH}$ is neglected in our analysis, but this gives a conservative constraints on the PBH parameters.
In other words, if we include the astrophysical contribution in our analysis,
the contribution of $\omega_{\rm PBH}$ allowed by the observation should be smaller compared 
to the present case and the excluded regions in Fig.~\ref{fig:2pcf_constrain} would be broadened.
If more high-$z$ quasars are discovered in the future, the measurement error will be reduced, and more accurate constraints will be obtained.

We also discussed a scenario where clumps of PBHs in early time collapses into SMBHs or its seeds.
Following the method of \cite{DeLuca:2022bjs}, we found that if the spectator field is very light ($c_I\ll1$), the creation of the seed of SMBHs at high-$z$ would be possible as shown in Fig.~\ref{fig:direct-collapse}.
However, the  correlation function in that  case is much larger than the one we derived from SHELLQs quasars at $z\sim6$, and hence such a scenario is also prohibited.

Finally we emphasize again that our result indicates that the PBH scenarios based on the spectator field 
as the origin of SMBHs at $z\sim6$ are almost excluded for a wide range of parameter space.
This may indicate that SMBHs are astrophysical origin unless an alternative cosmological model is suggested.

\section*{Acknowledgments}
We are grateful to Nobunari~Kashikawa and Junya~Arita for helpful and fruitful discussions. We also thank Taira Oogi for the communication on theoretical predictions of the angular correlation function for high-$z$ quasars. This work is supported in part by the MEXT KAKENHI Grant Number
17H06359~(TS), JP21H05453~(TS), and the JSPS KAKENHI Grant Number JP17H04830~(YM), 21H04494~(YM), 20H01949~(TN), 23H01215~(TN), JP19K03864~(TS) and 19K03874~(TT). 

The Hyper Suprime-Cam (HSC) collaboration includes the astronomical communities of Japan and Taiwan, and Princeton University. The HSC instrumentation and software were developed by NAOJ, the Kavli Institute for the Physics and Mathematics of the Universe (Kavli IPMU), the University of Tokyo, the High Energy Accelerator Research Organization (KEK), the Academia Sinica Institute for Astronomy and Astrophysics in Taiwan (ASIAA), and Princeton University. Funding was contributed by the FIRST program from Japanese Cabinet Office, MEXT, JSPS, Japan Science and Technology Agency (JST), the Toray Science Foundation, NAOJ, Kavli IPMU, KEK, ASIAA, and Princeton University. 

This paper is based on data collected at the Subaru Telescope and retrieved from the HSC data archive system, which is operated by Subaru Telescope and Astronomy Data Center (ADC) at NAOJ. Data analysis was in part carried out with the cooperation of Center for Computational Astrophysics (CfCA) at NAOJ. We are honored and grateful for the opportunity of observing the Universe from Maunakea, which has the cultural, historical and natural significance in Hawaii.

This paper makes use of software developed for Vera C. Rubin Observatory. We thank the Rubin Observatory for making their code available as free software at \url{http://pipelines.lsst.io/}.

\bibliography{ref}

\end{document}